\documentclass[a4paper,11pt]{article}
\pdfoutput=1 

\usepackage{jcappub} 

\usepackage[normalem]{ulem}
\usepackage{hyperref}
\usepackage{multirow}
\usepackage{array}

\usepackage{graphicx}
\usepackage{dcolumn}
\usepackage{bm}

 \usepackage{amssymb,amsmath}

 \usepackage{stackengine,graphicx}
 \usepackage[normalem]{ulem}

\title{Thermodynamics of 
$f(R)$ 
Theories of Gravity}

\author[a]{C.D. Peralta}
\author[b]{S.E. Jor\'as}

\affiliation[a]{
 Instituto de F\'isica, Universidad de Antioquia, Medell\'in, Colombia
}
 \affiliation[b]{
 Instituto de F\'\i sica, Universidade Federal do Rio de Janeiro,\\
 CEP 21941-972 Rio de Janeiro, RJ, Brazil
}

 \emailAdd{cesar.peralta@udea.edu.co}
 \emailAdd{joras@if.ufrj.br}

\abstract{
This paper starts from a toy model for inflation in a class of modified theories of gravity in the metric formalism. Instead of the standard procedure --- assuming a non-linear Lagrangian $f(R)$ in the Jordan frame --- we start from a simple $\phi^2$ potential in the Einstein frame and investigate the corresponding $f(R)$ in the former picture. The addition of an ad-hoc Cosmological Constant in the Einstein frame leads to a Thermodynamical interpretation of this physical system, which allows further insight on its (meta)stability and evolution.}

\begin{document}

\maketitle
\flushbottom

\section{Introduction}

Modified theories of gravity are ordinarily used to replace either the cosmological constant or the inflaton field -- explaining, respectively, the current and the early accelerated phase of expansion of the universe.
\textcolor{black}{Interestingly enough, only a handful of them are able to successfully describe (at the background level\footnote{The growth of cosmological perturbations is beyond the scope of the present paper.}) both of them \cite{2010JCAP...06..005A,2008PhRvD..77b6007N}, or a self-consistent inflationary model \cite{STAROBINSKY198099}, or even viable models of the present dark energy \cite{2007PhRvD..76f4004H,2007PhLB..654....7A,Starobinsky2007hu,2013PhRvD..88f3520O}.}
Nevertheless, \textcolor{black}{modifications on General Relativity (GR)} were introduced \cite{Bicknell_1974} long before any experimental data on either subject were available, just for the sake of completeness and diversity. 

In this paper we focus on $f(R)$ theories \cite{DeFelice:2010aj,Capozziello:2011et,Nojiri:2017ncd}  --- nonlinear functions of the Ricci scalar $R$ defined, as usual, in the Jordan Frame (JF). We follow the metric formalism, which features an extra degree of freedom (d.o.f), as we will briefly review.
It is well known that, upon a suitable conformal transformation (as we will also recall below), the modified gravitational Lagrangian assumes the usual Einstein-Hilbert form and the extra d.o.f. is materialized as a scalar field --- for obvious reasons, this is the so-called Einstein Frame (EF). \textcolor{black}{See, for instance, Ref.~\cite{2006PhLB..634...93C} for a discussion on how to determine the ``true physical frame''.}

Here, we will follow the same path, but in the opposite direction: we start from a standard $\phi^2$ potential with an {\it ad-hoc} Cosmological Constant $\Lambda$ in the EF (with also standard slow-roll initial conditions) and investigate the corresponding $f(R)$ in the JF. The introduction of  $\Lambda$ will  lead us to a full thermodynamical approach to $f(R)$ theories, shedding some light on the evolution of the system in both frames --- interesting results are still obtained even for the plain $\Lambda=0$ case.

We will now briefly review the aforementioned conformal transformation and the mapping from the quantities defined in one frame to their corresponding {\it Doppelg\"angers} in the other frame.

\section{Conformal Transformation and the Inverse Problem}
\label{conformal}

From now on, the super(sub)scripts ``$^E$" and  ``$^J$" indicate the frame (Einstein and Jordan, respectively) where the quantity is defined. We drop the subscript in $R_J\equiv R$ (and in $\phi_E\equiv \phi$ --- see below) to avoid excessive cluttering of the equations. 

We write the modified gravitation Lagrangian in JF (in the vacuum, i.e, no matter/radiation fields) as
\begin{equation}\label{Lvacuum}
L_J = \sqrt{-g^J} f(R),
\end{equation}
where $g^J\equiv \det (g^J_{\mu\nu})$. GR with a cosmological constant $\Lambda$ would correspond to $f(R) = R - 2\Lambda$. 
The standard variational procedure in the metric formalism yields fourth-order equations for the metric \cite{Sotiriou:2008rp}
\begin{equation}\label{EqfieldfR}
 R_{\mu\nu} f' -\frac12  g^J_{\mu\nu} f + g^J_{\mu\nu}\, \Box f' - \nabla_{\mu}\nabla_{\nu} f' = 0,
\end{equation}
where $f' \equiv {\rm d} f/{\rm d} R$. 

One then introduces the new pair of variables $\{g^E_{\mu\nu},p\}$, related to $g^J_{\mu\nu}$ (and to its derivatives) by a conformal transformation from the JF to the EF \textcolor{black}{\cite{Maeda, Magnano:1993bd,Wald:1984rg}}:
\begin{equation}
g^{E}_{\mu\nu} \equiv \Omega^2(x^\alpha) \, g^{J}_{\mu\nu}\, , \quad {\rm where} \quad  \Omega^2 \equiv p  \equiv f'(R).
\end{equation}
%
We now define $R(p)$ as a solution of the equation $f'[R(p)] - p = 0$. This procedure corresponds to a standard Legendre Transformation. As such, the expression $R(p)$ is uniquely defined as long as $f''\equiv d^2f/{\rm d} R^2$ has a definite sign. Nevertheless, it is possible to write a unique expression for $R(\phi)$ --- see Eq.~(\ref{Rphi}) below --- which holds across the branches where $f''(R)$ has different signs, and yields smooth functions $R(t)$ and $\phi(t)$ across the three branches. 

A scalar field  $\phi_E\equiv \phi$  (dropping the subscript) is traditionally defined in the EF by $p \equiv \exp{(\beta \,\phi)}$, with $\beta \equiv \sqrt{2/3}$. The Lagrangian (\ref{Lvacuum}) can then be recast in a more familiar form:
\begin{equation} \label{LagrangianE2}
L_E = \sqrt{-g^E} \Bigg[R_E - g_E^{\mu\nu} \phi_{,\mu} \phi_{,\nu} - 2V_E(\phi)\Bigg],
\end{equation}
where $R_E$ is the Ricci scalar obtained from $g^E_{\mu\nu}$. In other words, in the EF, the gravitational dynamics is set by a GR-like term ($R_E$) and the field $\phi$ is an ordinary minimally-coupled massive scalar field subject to the potential \cite{Magnano:1993bd}
\begin{equation}
V_E(\phi) \equiv \frac{1}{2p^2}\Big\{p(\phi) R[p(\phi)] - f[R(p(\phi))] \Big\}
\end{equation}
which is completely determined by the particular $f(R)$ chosen. 

In the present work we start by examining the inverse problem: from a scalar field $\phi$ and its potential $V_E(\phi)$,  we map $L_E$ in Eq.~(\ref{LagrangianE2}) onto the corresponding $L_J$ in Eq.~(\ref{Lvacuum}). 
Following a previoulsy established procedure \cite{Magnano:1993bd}, one arrives at the following parametric expressions:
\begin{align}
\label{fphi}
f(\phi) &= {\rm e}^{2 \beta \phi} \left[2V_E(\phi) + 2 \beta^{-1}  \frac{{\rm d} V_E(\phi)}{d\phi} \right] \quad {\rm and}\\
\label{Rphi}
R(\phi) &= {\rm e}^{\beta \phi} \left[4 V_E(\phi) + 2 \beta^{-1} \frac{{\rm d} V_E(\phi)}{d\phi} \right].
\end{align}
We will apply the above equations to the simplest possible (nontrivial) potential for a scalar field: $\sim \phi^2$. As we will mention further below, similar (but more intricate) results are obtained for more complex potentials. We allow a trivial shift ($a$) in the vacuum expectation value of the field which rescales the effective cosmological constant in the JF and yield cleaner plots. We also add an {\it ad hoc} Cosmological Constant $\Lambda$ and define then
\begin{equation}\label{VE}
V_E(\phi) \equiv \frac{1}{2}m^2_\phi \, (\phi-a)^2 + \Lambda.
\end{equation}
One might argue that the insertion of $\Lambda$ goes completely against the reasoning of modifying GR but, for now, $\Lambda$ is written just for the sake of completeness. As we will see later on, it will turn out to be a key ingredient for the thermodynamic interpretation. Nevertheless, even the standard case where $\Lambda=0$ yields very interesting results.

\begin{figure}
\center
\includegraphics[width=0.47\textwidth]{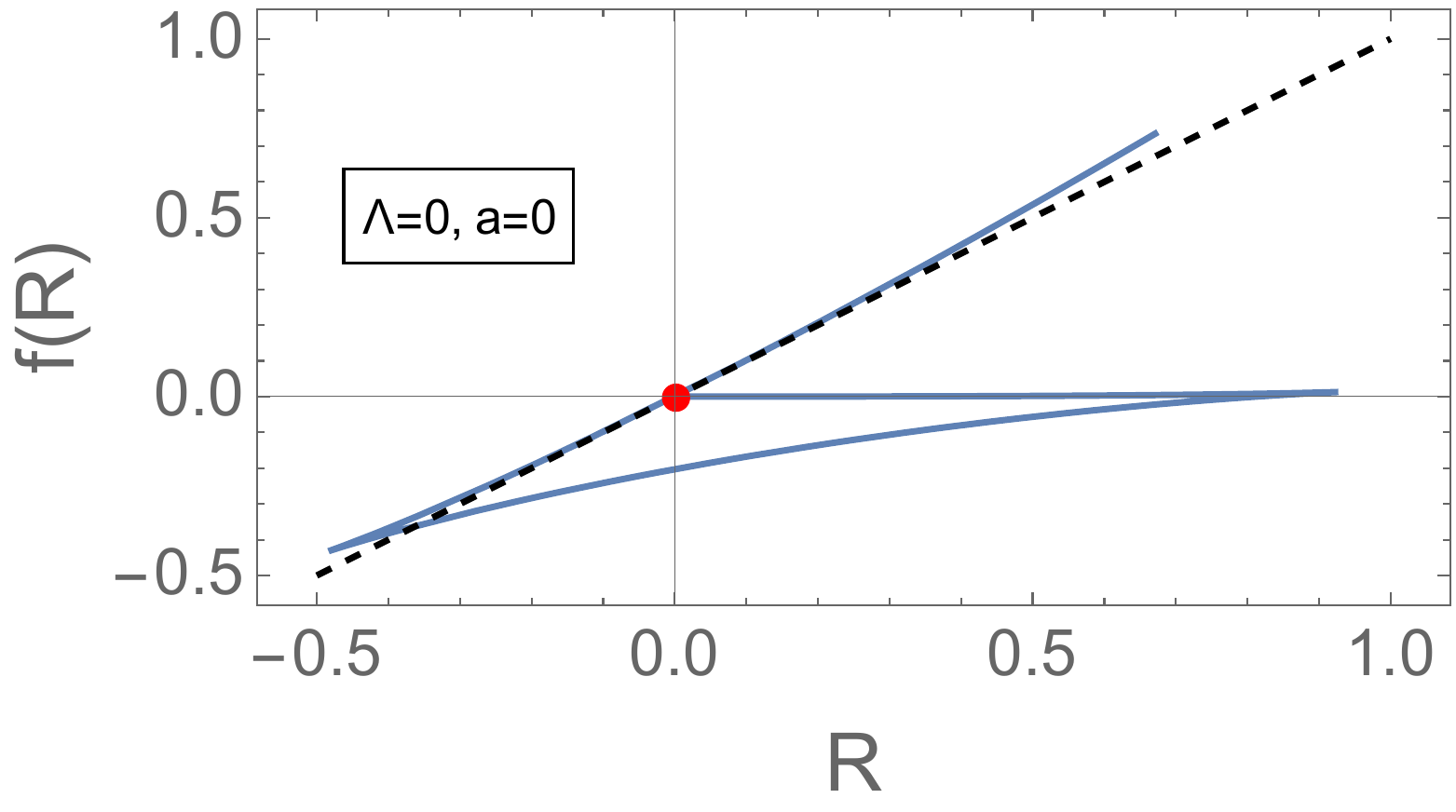}
\includegraphics[width=0.47\textwidth]{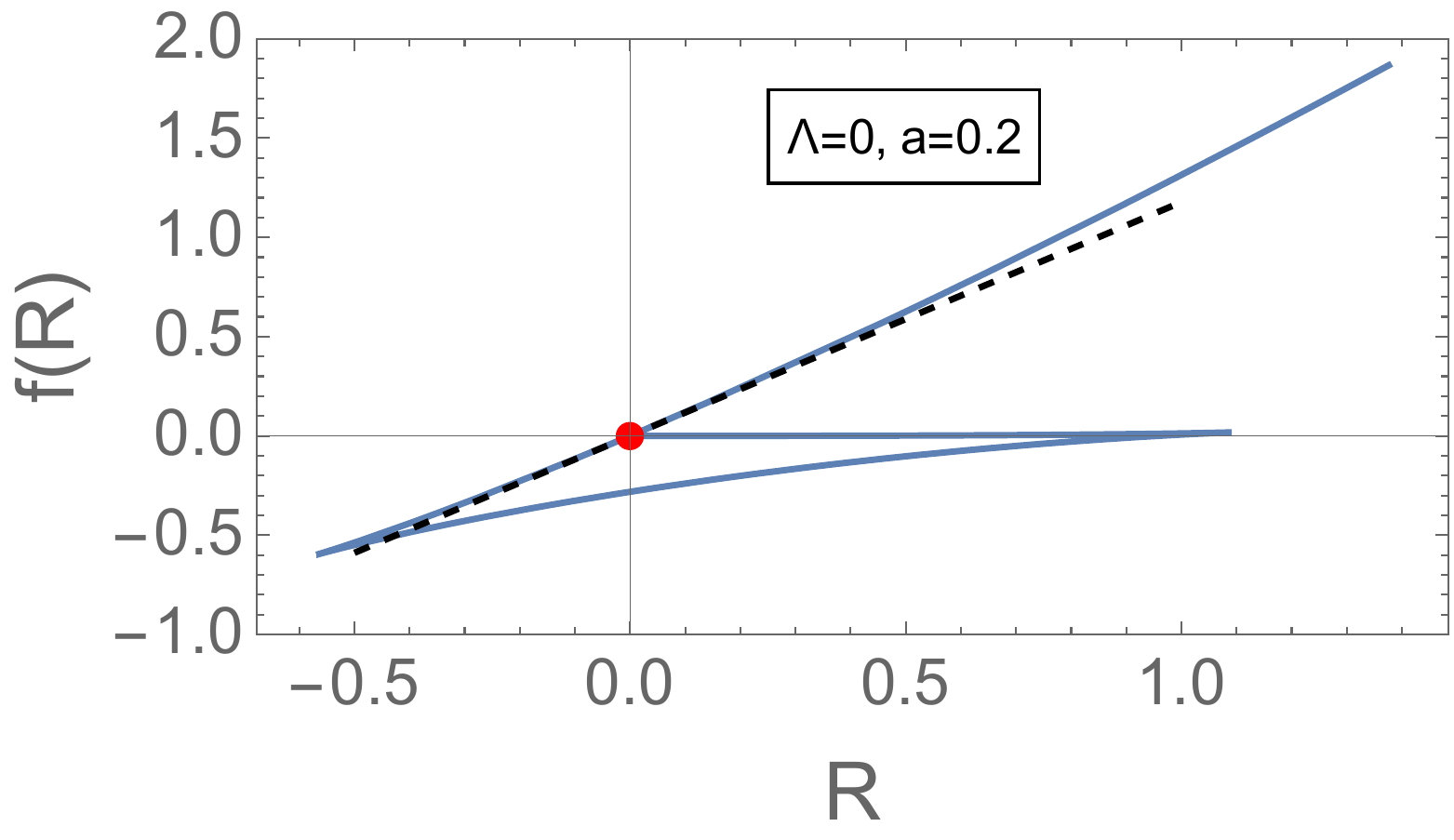}
\includegraphics[width=0.45\textwidth]{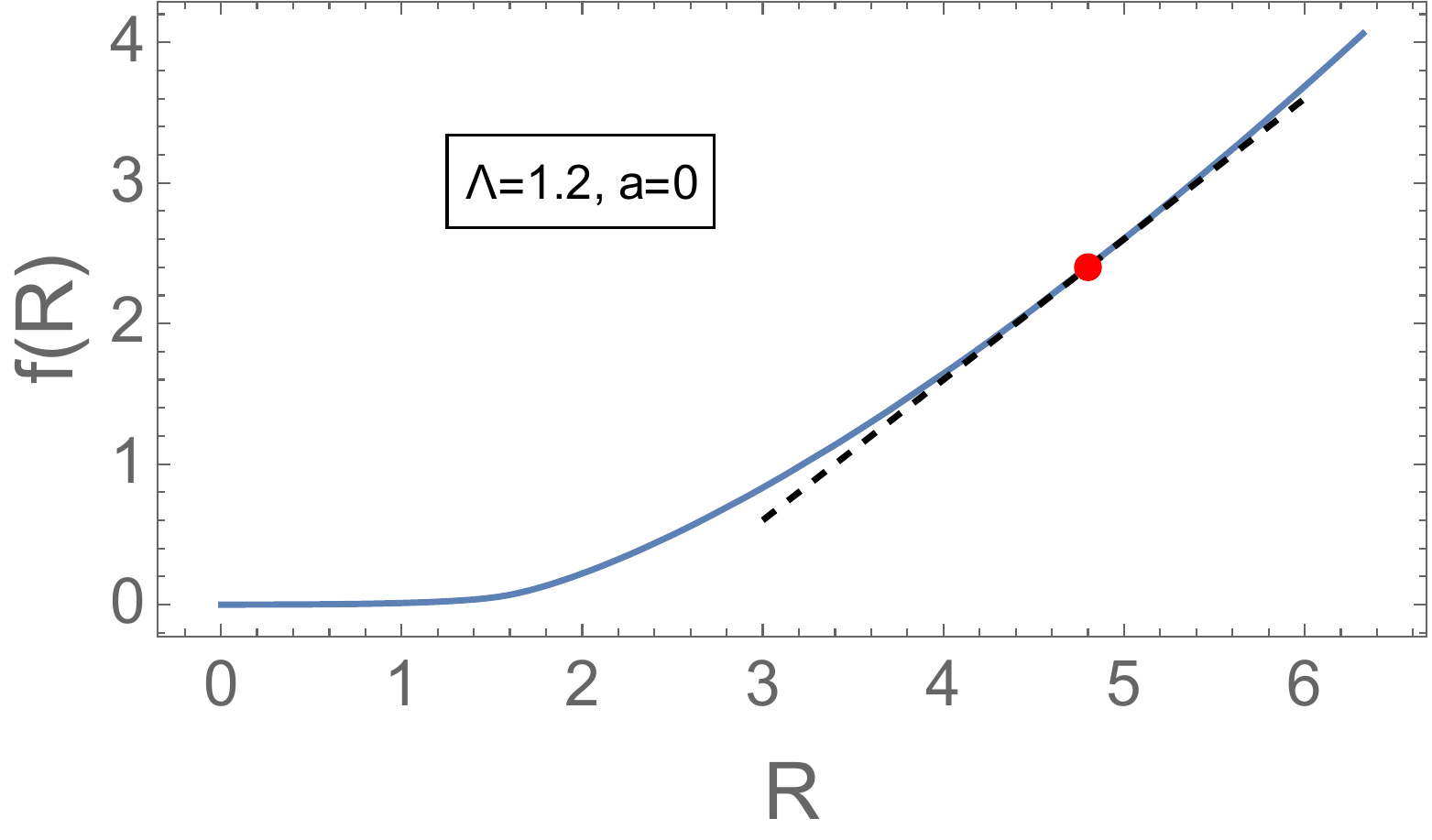}
\includegraphics[width=0.45\textwidth]{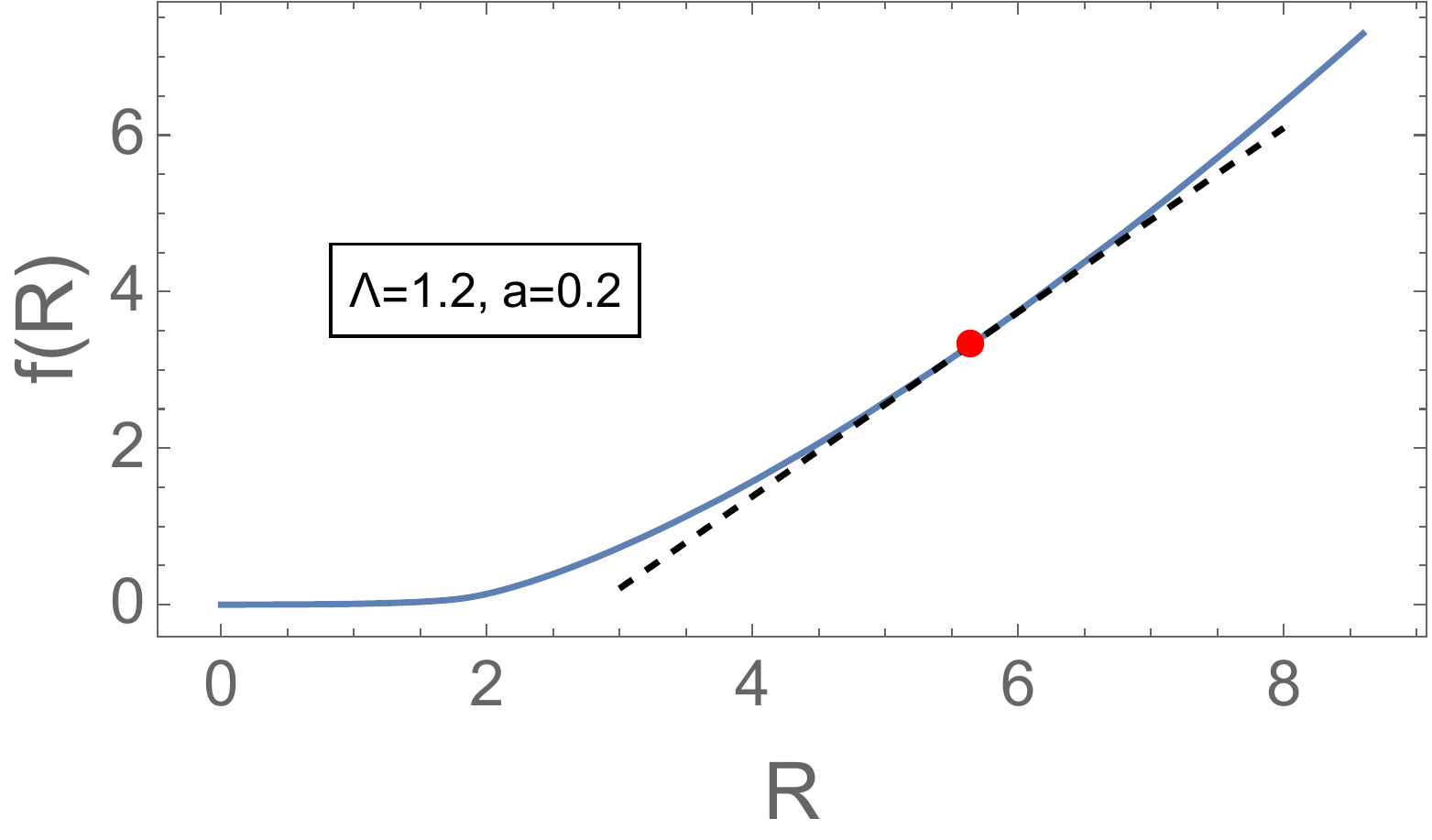}
\caption{Parametric plots of $f(R)$ given by Eqs. (\ref{fVE},\ref{RVE}) for $\phi \in [-15,0.5]$ and for the parameters $\{\Lambda,a\}$ shown in the respective insets. In all panels, $f'>0 \, \forall R$.  
Note that the panels with low $\Lambda$ present a 3-branch structure. In all of them, the dashed line is given by $f(R) = \exp(\beta a) R - 2 \Lambda_J$, the linear behavior of $f$ at the red dot, which  indicates the final de Sitter solution (Minkowski, if $\Lambda=0$), reached when $\phi=a$. The field $\phi$ and $a$ are given in Planck-Mass ($M_{\rm pl}$) units, $R$ and $\Lambda$ are given in $M_{\rm pl}^4$. We used $m_\phi=1 M_{\rm pl}$.
}
\label{swallowtail}
\end{figure}

We then obtain the corresponding parametric form of $f(R)$:
\begin{align}
\label{fVE}
f(\phi) &= e^{2 \beta  \phi } \bigg[m_\phi^2 (a-\phi ) \Big(a  -  \phi -\frac{2}{\beta}\Big)+2 \Lambda \bigg]\\
\label{RVE}
R(\phi) &= 2 e^{\beta  \phi } \bigg[ m_\phi^2 (a-\phi ) \Big(a  -  \phi -\frac{1}{\beta}\Big)+2 \Lambda \bigg],
\end{align}
which we plot in Fig.~\ref{swallowtail}. If $\Lambda<\Lambda_c$ (to be defined later on), the curve features a 3-branch structure. Throughout the paper, we will refer to those three stages as {\it branches} of the system. 
In all of them, from the above expressions, one has $df/dR\equiv f' = \exp(\beta \phi)>0$ as usually required \cite{DeFelice:2010aj, Magnano:1993bd, Starobinsky:2007hu} for non-repulsive gravity. In particular, on the final branch, when the field $\phi$ oscillates around its potential minimum ($\phi=a$), one recovers GR only if $f'=\exp(\beta a)=1$, i.e, if $a=0$.
If $\Lambda \neq 0$ and regardless of $a$, the system does reach a de Sitter state with a corresponding effective cosmological constant in the JF, given by $\Lambda_J  \equiv \Lambda \exp(2 \beta a)$, a non vanishing $R= R_{\rm dS}\equiv 4 \Lambda \exp(\beta a)$ and an effective gravitational constant  $G_{\rm eff}\equiv G_N \exp(-\beta a)$, where $G_N$ is the standard gravitational constant. In other words, at the final stage ($\phi \approx a$), the modified Lagrangian given by Eqs.~(\ref{fVE}) and (\ref{RVE}) can be written as the linear function
$f(R) = \exp(\beta a) R - 2 \Lambda_J$.
Such swallowtail-like structure has already been pointed out \cite{MIELKE_2008}, but with a different interpretation of the effective gravitational and cosmological  constants --- we define them only in the final stage (when the field settles down), as opposed to the intermediate, unstable phase.

\begin{figure}
\center
\includegraphics[width=0.42\textwidth]{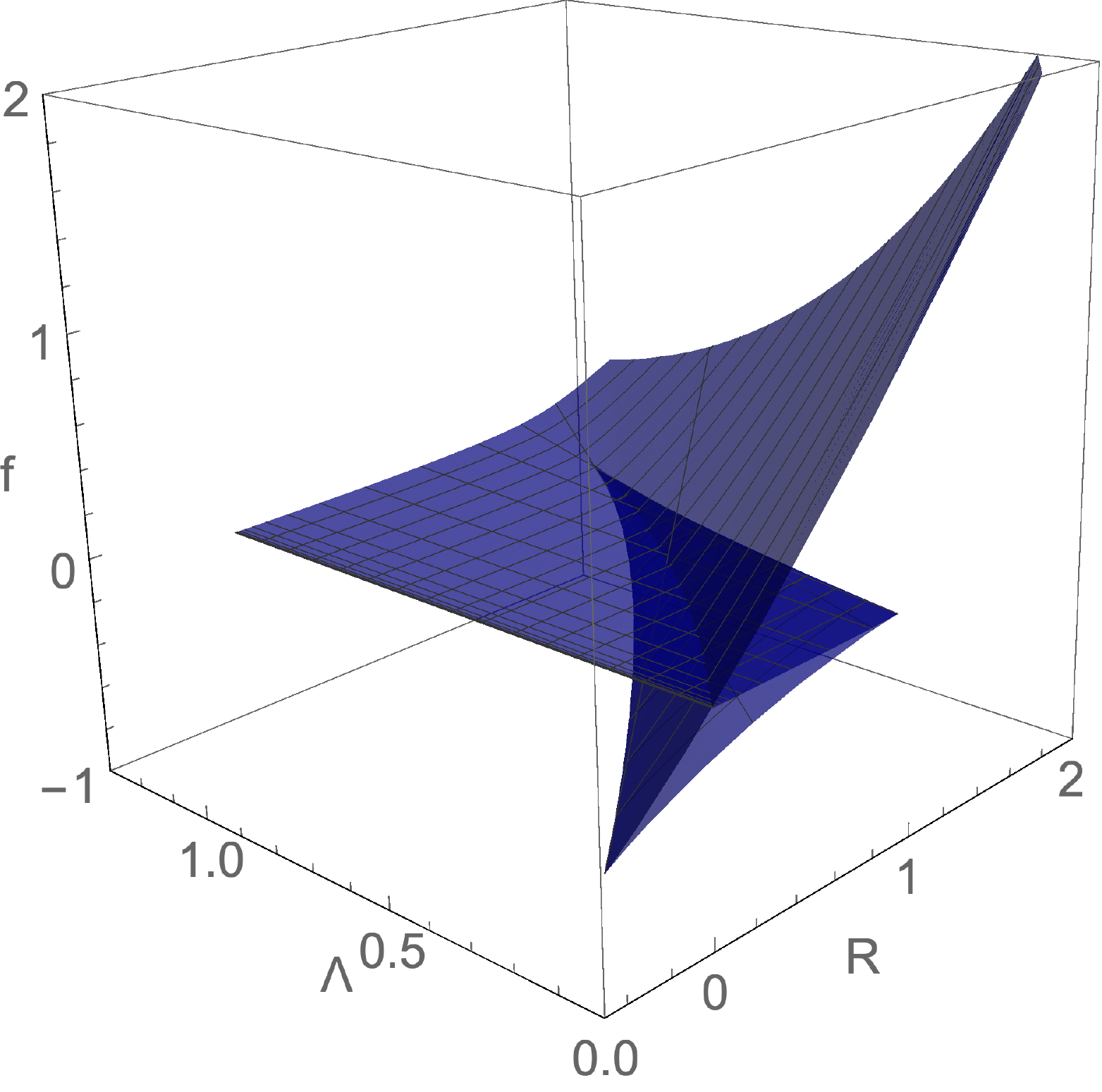}
\includegraphics[width=0.42\textwidth]{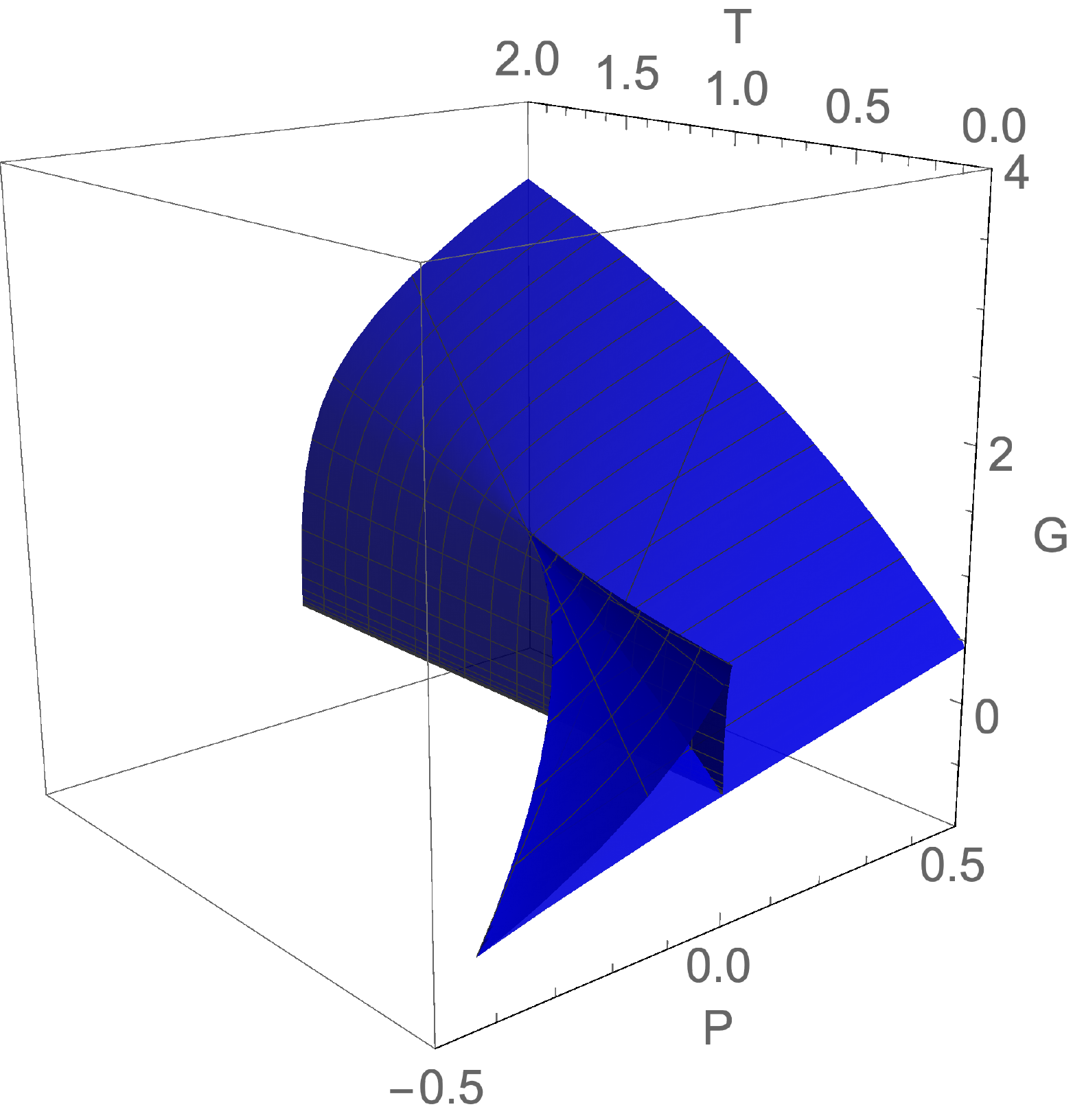}
\caption{Plots of $f(R,\Lambda)$, given by Eqs.~(\ref{fVE}) and (\ref{RVE}), and $G(P,T)$, given by Eqs.~(\ref{Gphi}) and (\ref{Pphi}) with $\beta=\sqrt{2/3}$ and $a=0$. The latter panel can be accessed on line (for a turnable picture) at \url{https://tinyurl.com/s7am2px}}
\label{swallow3D}
\end{figure}

The behaviour of $f(R)$ for different values of $\Lambda$ is shown in Fig.~\ref{swallow3D} --- two slices with $\Lambda=0$  and $\Lambda=1.2$ are shown in Fig.~\ref{swallowtail} (left panels). The attentive reader may recognize a similiar surface for the van der Waals gas \cite{callen} (vdW, from now on). Indeed, Fig.~\ref{swallow3D} bears strong resemblance to the Gibbs potential $G$ for the vdW gas as a function of its temperature $T$ and its pressure $P$. More remarkable is the region where $f''<0$, which indicates a tachyonic instability
\cite{Sotiriou:2008rp}: here, it corresponds to the upper curve of the standard $G(P,T)$ curve, which is also unstable.

\section{Thermodynamics}

The mere similarity between Fig.~\ref{swallow3D} and the Gibbs potential might be just a coincidence. Nevertheless, there is indeed a deepe{black}r connection: the whole system --- its equilibrium points, stability and evolution --- is determined by the internal energy $U$, the Gibbs potential $G$ and its critical points, as we will now see. 

For now, let us associate the Cosmological Constant $\Lambda$ to an effective temperature $T\equiv\Lambda$. 
\textcolor{black}{It is well known \cite{Figari,Gibbons}  that, in a de Sitter-like spacetime, a cosmological constant corresponds to an effective temperature due to the presence of the horizon, just like for a black hole. Thus, such correspondence comes with no surprise.}
On the other hand, we do not directly identify $G$ to $f$ and neither $P$ to $R$. We rather use an slightly more general {\it Ansatz}: we define a new pair of coordinates $\{-G,P\}$ as a rotation of the original one $\{f,R\}$:
\begin{equation}
\left(
\begin{tabular}{c}
$-G$ \\
$P$
\end{tabular}
\right)
\equiv 
\left(
\begin{tabular}{cc}
$\cos\theta$ &  $- \sin\theta$ \\
$\sin\theta$ & $\cos\theta$
\end{tabular}
\right)
\left(
\begin{tabular}{c}
$f$ \\
$R$
\end{tabular}
\right).
\end{equation}
which yields
\begin{align}
 G (\phi,T) &=
 e^{\beta  \phi } \sin (\theta ) \left(\frac{2 (\phi -a) (\beta  (\phi -a)+1)}{\beta }+4 T\right) + \nonumber \\
  &-e^{2 \beta  \phi } \cos (\theta ) \left(\frac{2 (\phi -a)}{\beta }+(\phi -a)^2+2 T\right)
 \label{Gphitheta}  \\
P (\phi,T) &=e^{2 \beta  \phi } \sin (\theta ) \left(\frac{2 (\phi -a)}{\beta }+(\phi -a)^2+2 T\right)+\nonumber \\
&+e^{\beta  \phi } \cos (\theta ) \left(\frac{2 (\phi -a) (\beta  (\phi -a)-1)}{\beta }+4 T\right)
\label{Pphitheta}
\end{align}

The effective volume $V$ is the variable ``canonically conjugated" to the effective pressure $P$, i.e, since
\begin{equation}
dG(P,T) = V \cdot  dP - S \cdot dT,
\label{dG}
\end{equation}
one can define an effective volume
\begin{equation}
V \equiv \left.\frac{\partial G}{\partial P}\right|_T = \left.\frac{\partial G/\partial\phi}{\partial P/\partial\phi}\right|_T = \frac{1-e^{\beta  \phi } \cot (\theta )}{e^{\beta  \phi }+\cot (\theta )},
\label{Vphitheta}
\end{equation}
which can be inverted and yield
\begin{equation}
\phi = \frac{1}{\beta}\log \left(\frac{1-V \cot (\theta )}{\cot (\theta )+V}\right).
\end{equation}

\begin{figure}[t]
\center
\includegraphics[width=0.4\textwidth]{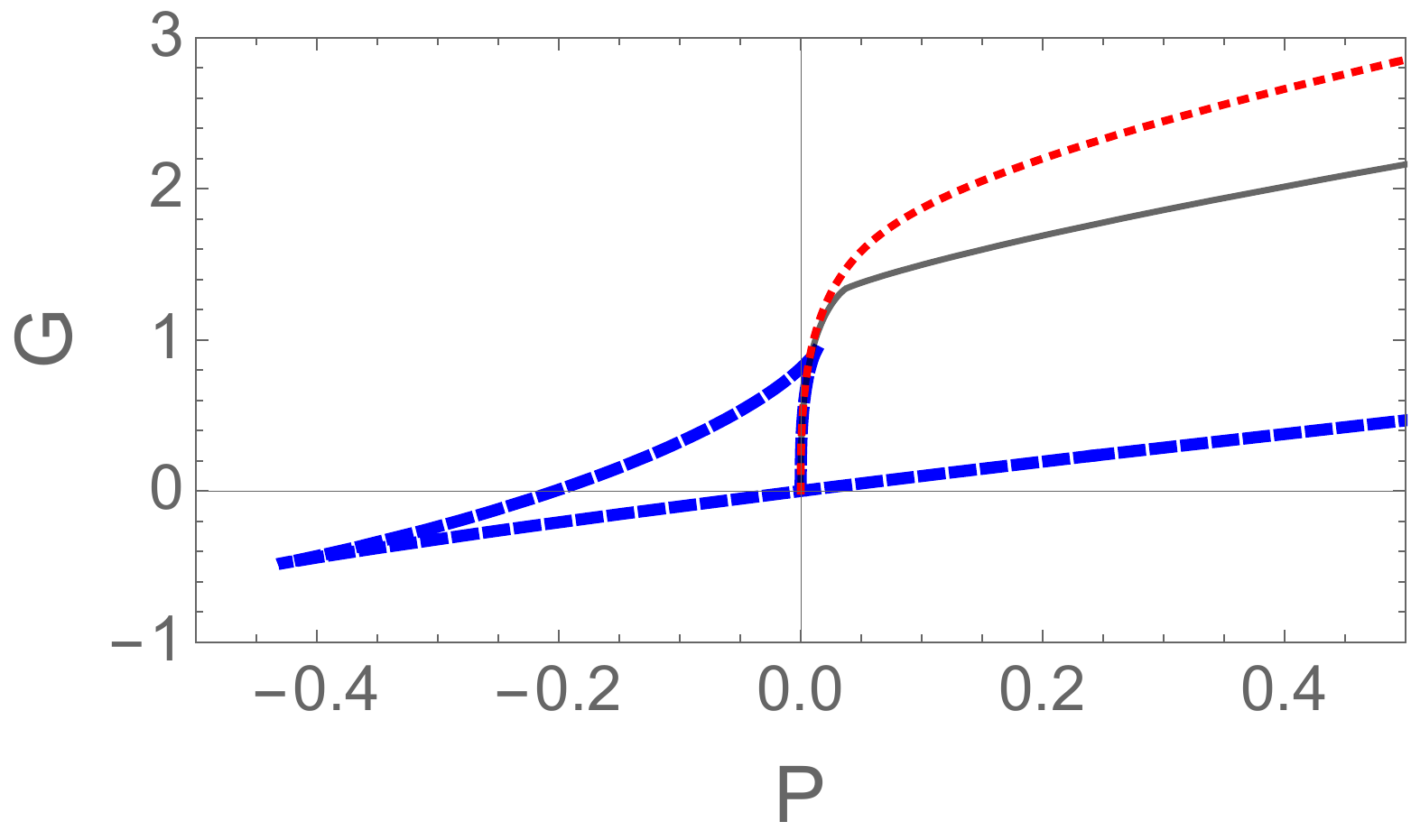}
\caption{Plot of the Gibbs Potential $G$ as a function of the pressure $P$, for $\beta=\sqrt{2/3}$, $\theta=\theta_*$ and $T=0$ (dashed blue), $T=T_c=15/16$ (solid black) and $T=1.5$ (dotted red). }
\label{Gphi_fig}
\end{figure}

In order to define the exact correspondence, i.e, the value of $\theta$, we only require that the volume  is positive and unlimited from below.
Indeed, such procedure yields $\theta = \theta_* \equiv  \pi/2$ and simpler parametric expressions for the previously defined thermodynamic quantities: 
\begin{align}
G &= e^{\beta  \phi } \left(\frac{2 (\phi -a) [\beta  (\phi -a)+1]}{\beta }+4 T\right) \label{Gphi}\\
P &= e^{2 \beta  \phi } \left(\frac{2 (\phi -a)}{\beta }+(\phi -a)^2+2 T\right) 
\label{Pphi}\\
V &= \exp(-\beta \phi) \quad \Leftrightarrow \quad \phi = -\frac{1}{\beta}\log(V).
\label{Vphi}
\end{align}
and the corresponding  plot in Fig.~\ref{swallow3D} (right panel). In Fig.~\ref{Gphi_fig}, we plot the curve $G(P)$ for different temperatures $T$, each one corresponding to a different vertical section of the previous 3D figure. 

One can also calculate the Helmholtz energy 
\begin{align}
F(T,V) &\equiv G - P \cdot V   = \frac{e^{2 \beta  \phi } \csc (\theta ) \big[(a-\phi )^2+2 T\big]}{e^{\beta  \phi }+\cot (\theta )} \\
& = \frac{1}{V}\bigg[\left(a+\frac{1}{\beta }\log V\right)^2+2 T\bigg] \qquad {\rm if} \quad \theta=\pi/2
\end{align}
from which one can define the entropy as
\begin{align}
S (T,V) &\equiv - \left.\frac{\partial F}{\partial T}\right|_V = -\frac{2 \sin (\theta ) (V \cot (\theta )-1)^2}{\cot (\theta )+V} \\
& = -\frac{2}{V}  \qquad {\rm if} \quad \theta=\pi/2
\end{align}
One can then realize that the specific heat at constant volume vanishes, since $C_V \equiv T \cdot \partial S/\partial T|_V  = 0 \, \forall T$. Such feature is not unusual: it has been already found in studies of  thermodynamics and phase transitions of black holes \cite{Dolan_2011}.

\begin{figure}[t]
\center
\includegraphics[width=0.4\textwidth]{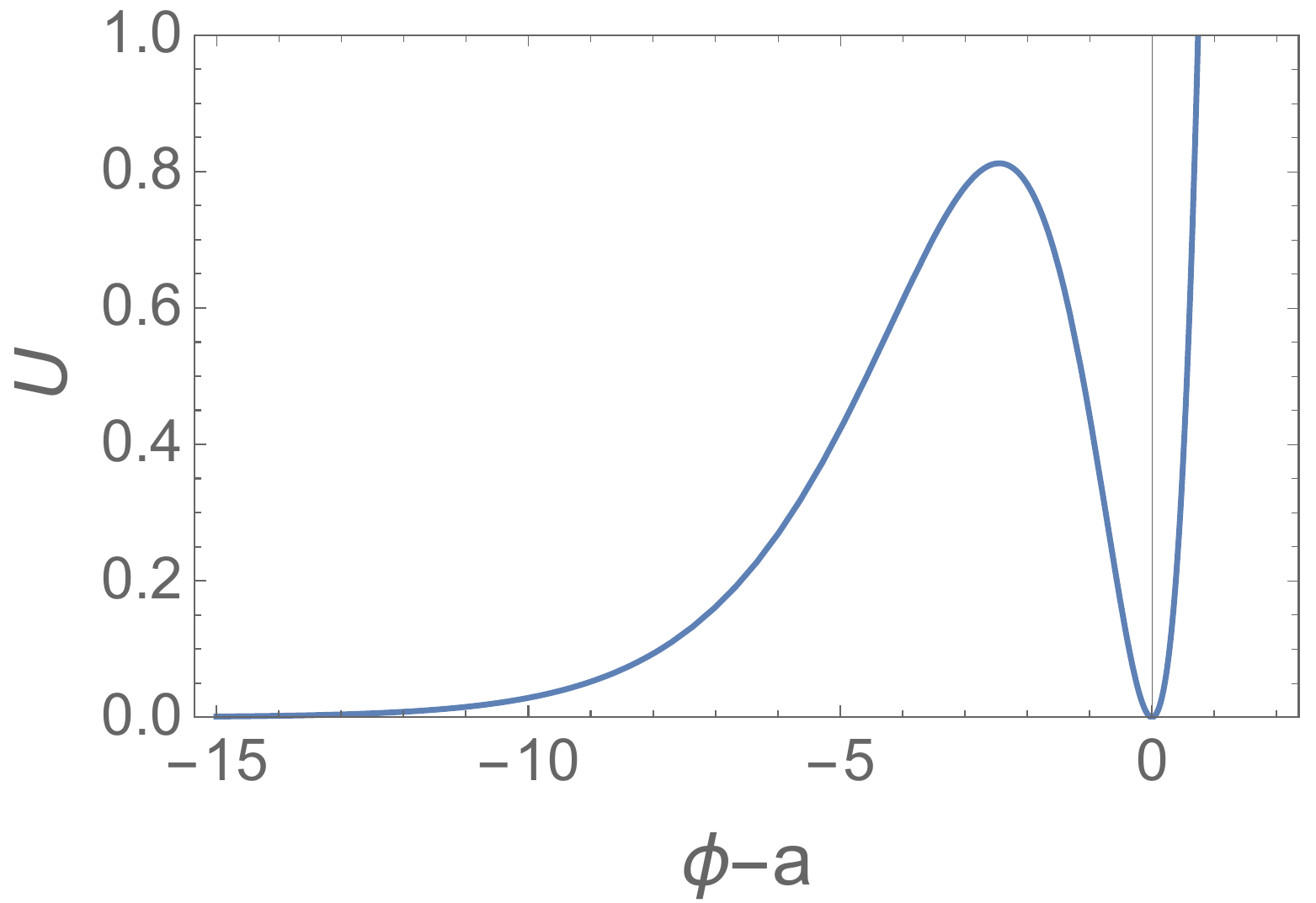}
\includegraphics[width=0.4\textwidth]{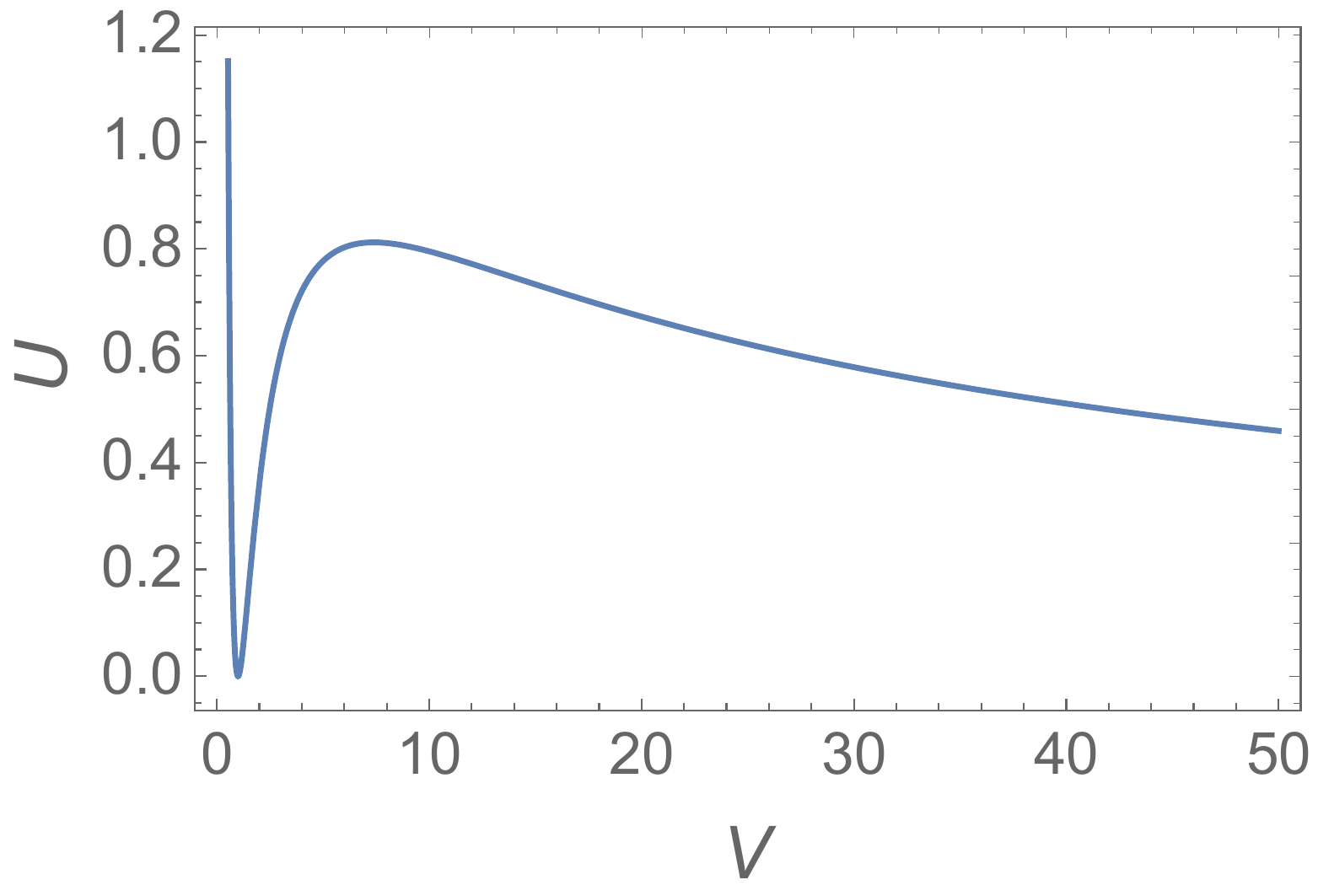}
\caption{Plot of the \textcolor{black}{internal energy} $U$ with $\theta=\theta_*=\pi/2$ as a function of $\phi-a$ (left panel) and of the volume $V$ with $a=0$ (right panel). We recall that $V = \exp(-\beta \phi)$. }
\label{Ufigs}
\end{figure}

The internal energy $U(T,V)$ is given by its standard definition:
\begin{align}
U &\equiv G - P \cdot V + T \cdot S = \frac{(a-\phi )^2 e^{2 \beta  \phi } \csc (\theta )}{e^{\beta  \phi }+\cot (\theta )}  \\
&= \frac{1}{V} \left(a+\frac{1}{\beta }\log V\right)^2 \qquad {\rm if} \quad \theta=\pi/2
\end{align}
for which $\phi=a$ (accordingly, $V=\exp(-\beta a)$) is always a minimum. It turns out that also $U$ is only a function of the volume $V$ and {\it not} of the temperature $T$. One might acknowledge the existence of another two equilibrium points: an asymptotic one (a local minimum at $\phi\to-\infty$) and a local maximum (whose position depends on $\theta$) --- see Fig.~\ref{Ufigs}.

\begin{figure}[t]
\center
\includegraphics[width=0.7\textwidth]{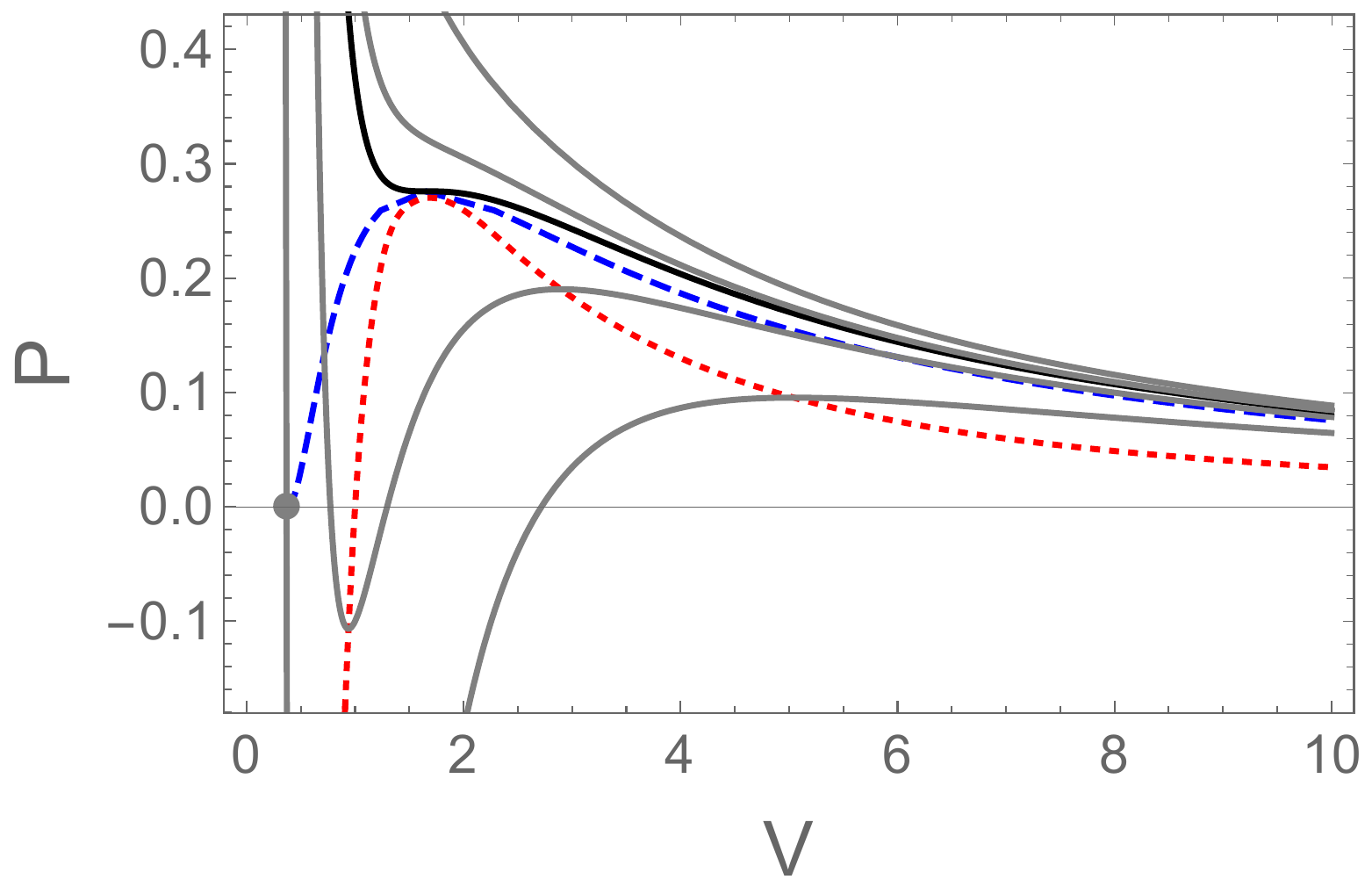}\\ 
\hspace{1cm}
\includegraphics[width=0.3\textwidth]{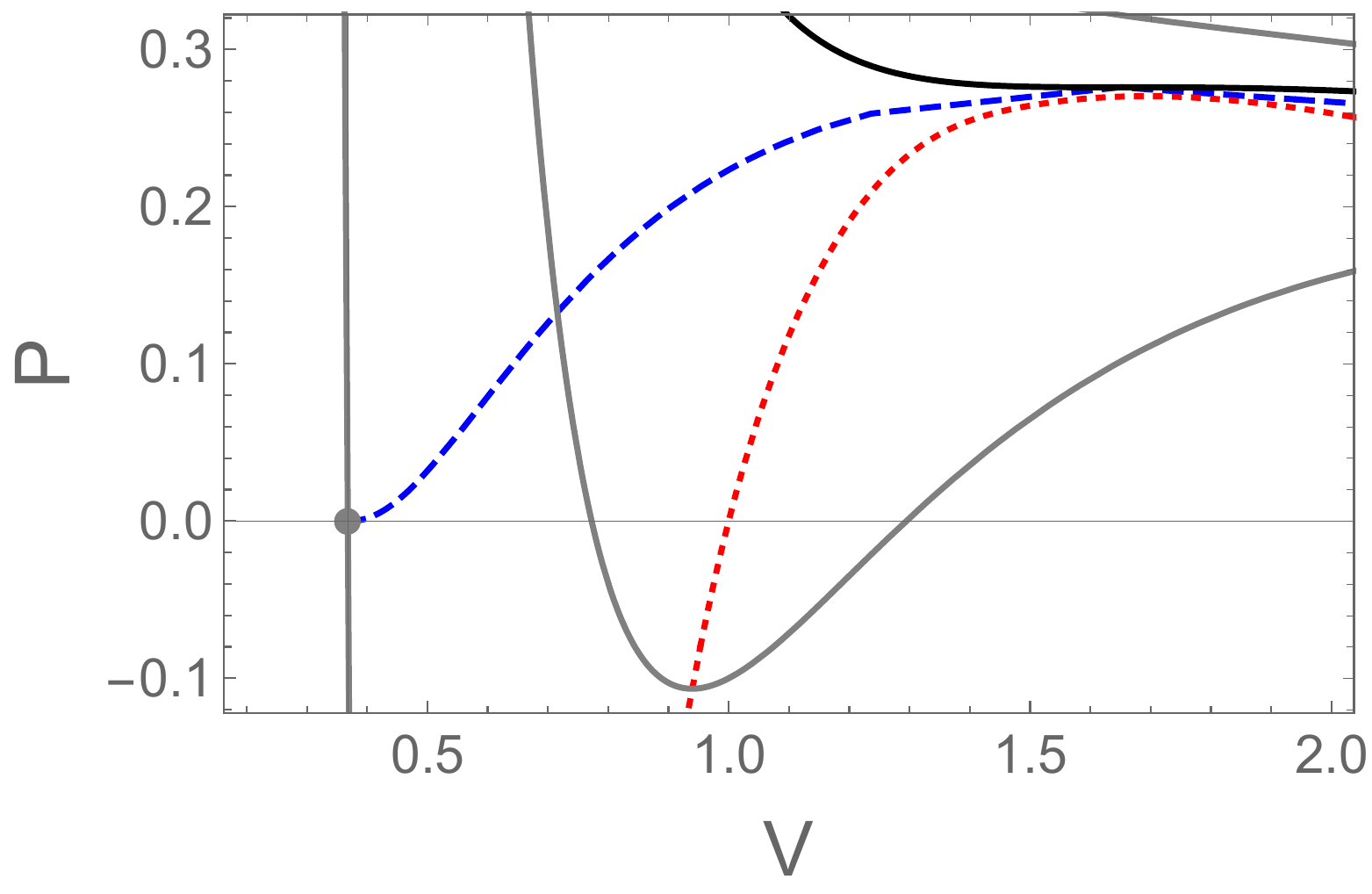}
\includegraphics[width=0.3\textwidth]{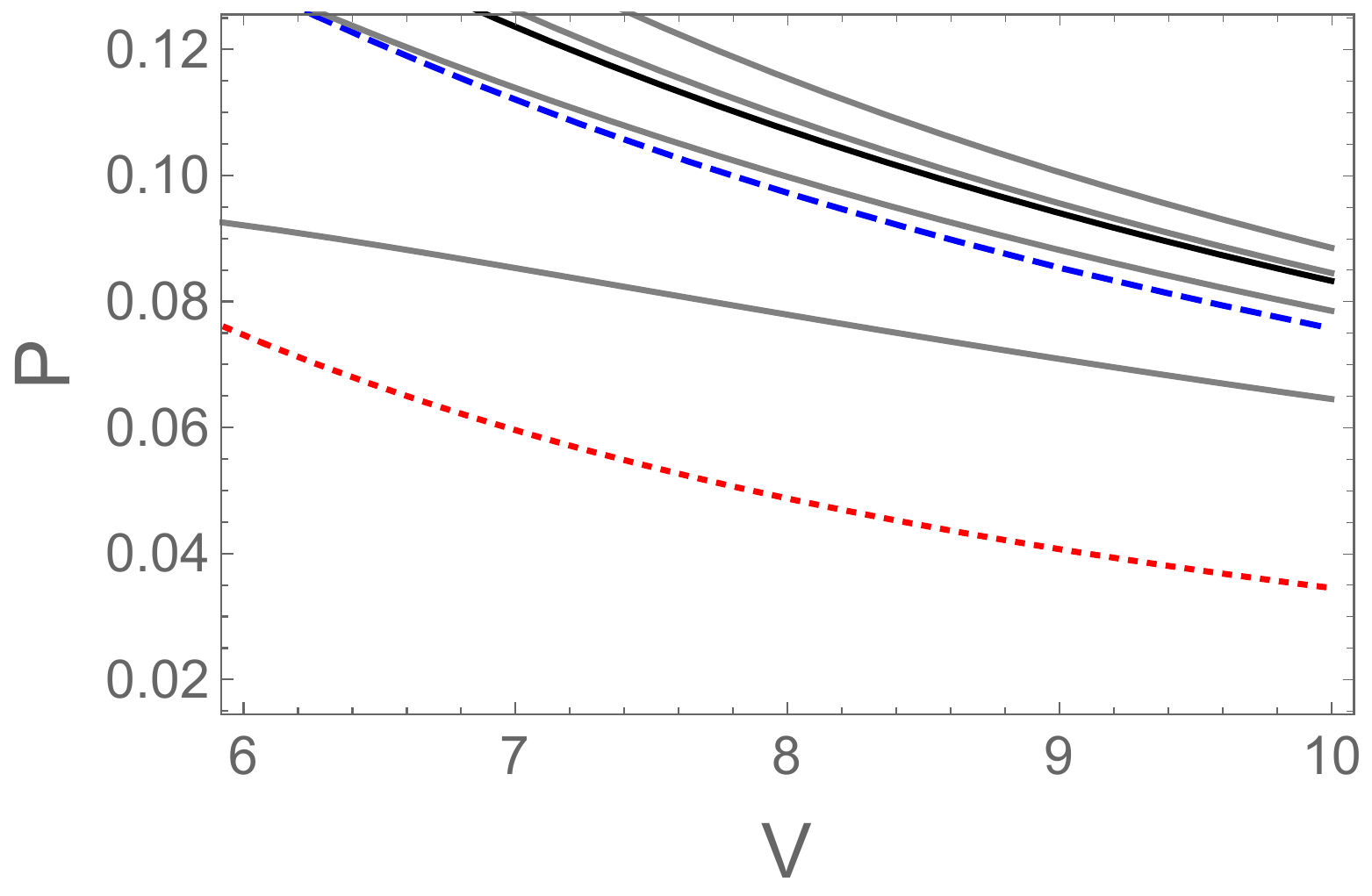}
\caption{Plot of the effective pressure $P$ as a function of the effective volume $V$, for $a=a_*\equiv 1/\beta$ and different values of temperature: $T=T_c\equiv 15/16$ (solid thick black); lower (higher) curves, in solid thin gray, correspond to lower (higher) temperatures. The {\bf spinodal curve} is plotted in dotted red. The {\bf binodal curve} is plotted in dashed blue. The gray circle indicates the final configuration ($\phi=a$) for the $T=0$ case (higher temperatures correspond to  higher final pressures). {\bf Lower panels:} zooming into the left and right-hand ends of the volume axis to show the behavior of the same curves. Note that the binodal curve does end at the gray circle.}
\label{sbin}
\end{figure}

 From now on, we shall always use $\theta=\theta_*=\pi/2$.
Equations~(\ref{Pphi}) and (\ref{Vphi}) yield the equation of state for our vdW-like  ``efective gas'', i.e, an expression that relates $P$, $V$ and $T$:
\begin{equation}
P = \frac{\beta  \left(a^2 \beta -2 a+2 \beta  T\right)+(2 a \beta -2 +\log V )\log V }{\beta ^2 V^2}.
\label{PV}
\end{equation}
The behaviour of $P(V)$ for four different values of $T$ is shown in Fig.~\ref{sbin}, which bears strong resemblance to a vdW gas\footnote{Nevertheless, here one obtains $P\propto T V^{-2}$ in the high-temperature limit, instead of the standard ideal-gas behavior $P \propto T V^{-1}$.}. Even though the equations of state are not exactly the same, they do describe the same phenomena, as we will now see. 

For instance, we can define the binodal and spinodal curves, that indicate, respectively, the regions of metastability and instability of the system --- see Fig.~\ref{sbin}. The former can be obtained using two equivalent calculations --- from the self-intersecting points of the Gibbs function and from the Maxwell construction --- supporting the results from each other. The latter curve is obtained from the extrema of the Gibbs function (see Fig.~\ref{swallowtail}), i.e, the first two turning points (extrema) of $R(t)$. The {\it critical point} $\{P_c,T_c,V_c\}$, defined at the crossing of those curves, indicates the end of the coexistence line. We will come back to those curves in the next section.

\begin{figure}[t]
\center
\includegraphics[width=0.4\textwidth]{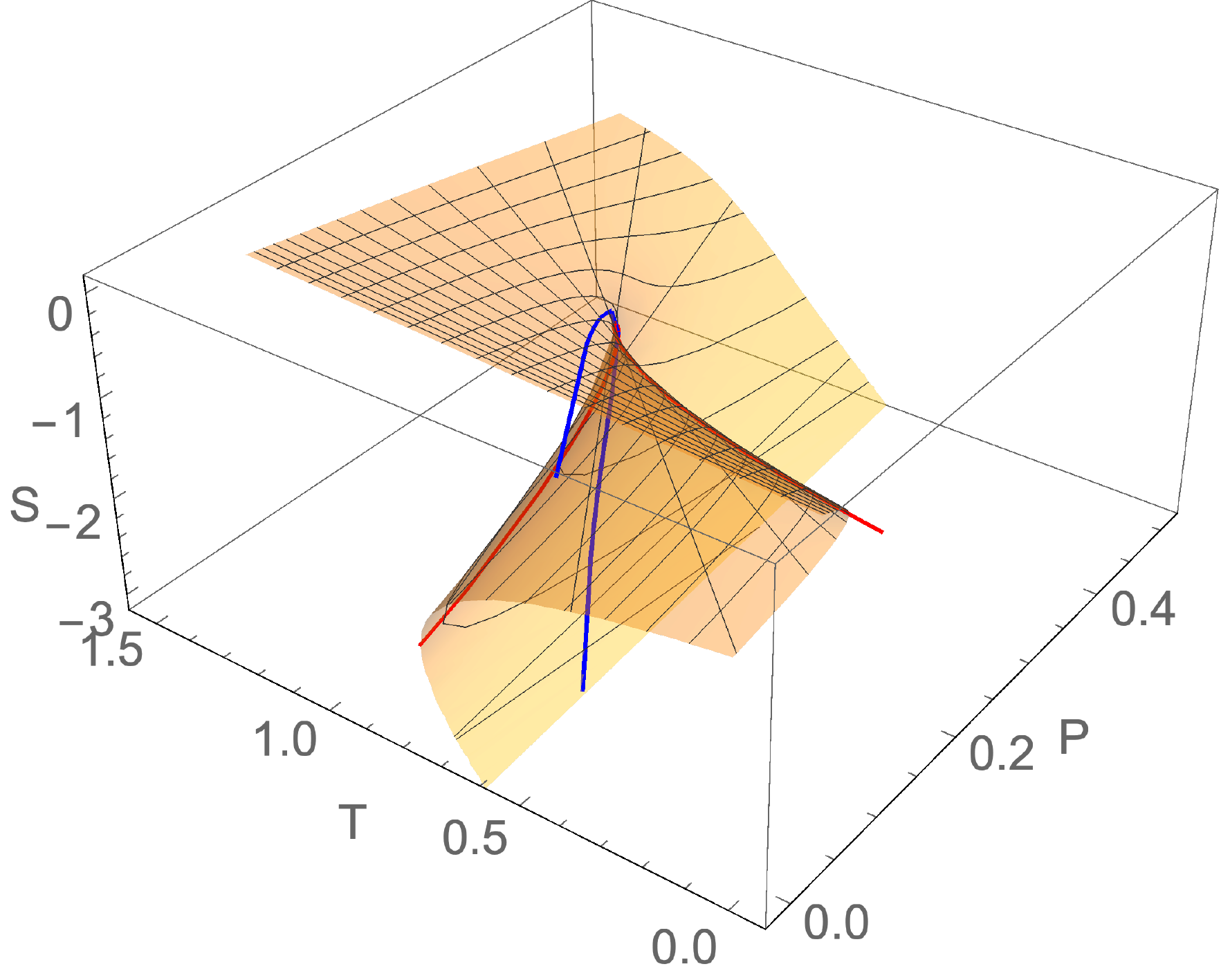}
\caption{Surface given by $S(P,T)$ for $a=a_*$.  The spinodal and binodal curves are indicated in red (horizontal cusp shape) and blue (vertical ``$\subset$" shape), respectively. A turnable version is available at \url{https://tinyurl.com/wopt7fq}.}
\label{spt}
\end{figure}

The entropy as a function of pressure and temperature provides another very important piece of information.  $S(P,T)$ is depicted in Fig.~\ref{spt}, which also shows the spinodal and binodal curves.  The region where the entropy is multi-valued is known in Catastrophe Theory \cite{saunders1980introduction} as a cusp and indicates the existence of a first-order phase transition and unstable configurations.

\begin{figure}[t]
\center
\includegraphics[width=0.4\textwidth]{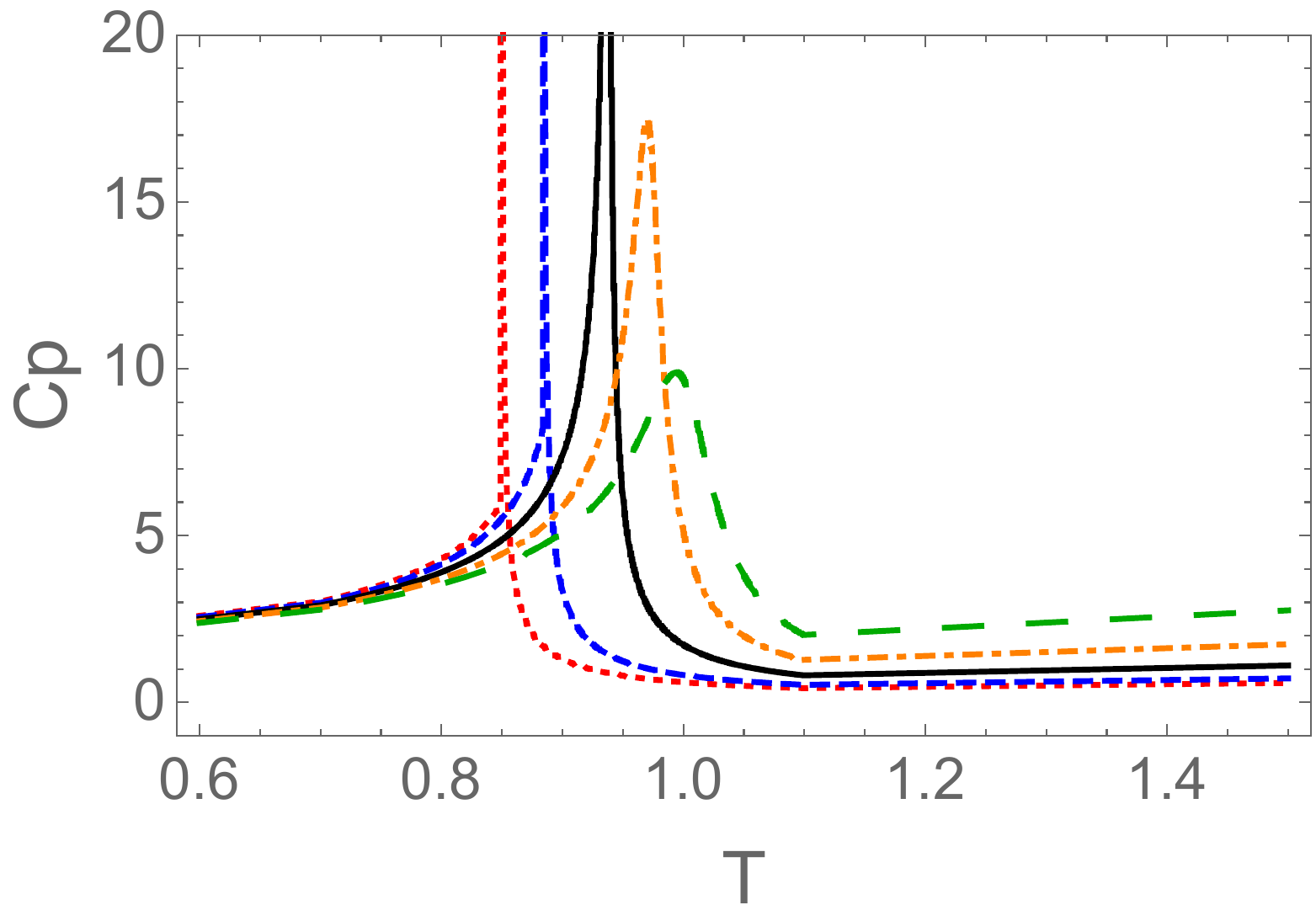}
\caption{Behavior of the specific heat at constant pressure $C_P$ as a function of the temperature $T$ close to its transition value ($T_c=15/16\approx 0.94$ if $P=P_c$), for different values of pressure (from left to right): $0.85 P_c$ (dotted red), $0.85 P_c$ (dashed blue), $P_c$ (solid black), $1.1 P_c$ (dot-dashed orange) and $1.2 P_c$ (long-dashed green). In all curves, $a=a_*$, for which $P_c\approx 1.51$.}
\label{Cp}
\end{figure}

From $S(P,T)$ we can get the specific heat at constant pressure, $ C_P \equiv T \cdot \partial S/\partial T |_P$, shown in Fig.~\ref{Cp}. We obtain the expected behavior for temperatures around the coexistence curve, for pressures both below (finite jump) and above (smooth behavior) the critical value $P_c$. We also obtain the usual divergence at the critical point $\{T_c,P_c\}$ (solid black line in Fig.~\ref{Cp}) as given by $C_P|_{P_c} \sim [(T_c-T)/T_c]^\alpha$, with $\alpha\approx 1.00$. 

\section{A Numerical Example}

From now on, we will investigate the potential given in Eq.~(\ref{VE}) as a standard toy-model inflationary potential in the EF --- initially, we will keep $a=\Lambda=0$, except when necessary for a cleaner picture and noted so. 

First of all, we have to determine the time evolution of $R(t)$ and $\phi(t)$. We recall that throughout this paper there is no matter nor radiation; the $\phi$ field is pure gravity. In GR, that would imply $R=0 \, \forall \, t$. In $f(R)$ theories, on the other hand, $R$ has a dynamical behavior of its own. Here, it suffices to use $R[\phi(t)]$ (defined in the JF) from Eq.~(\ref{RVE}) and  $\phi(t)$ (in the EF) from the standard equation of motion for a scalar field in an expanding homogeneous spacetime:
\begin{equation}
 \ddot{\phi}(t) + 3 H(t) \dot{\phi}(t) +  V_E'[\phi(t)]=0,
 \label{eqphi}
\end{equation}
where $V_E'\equiv dV_E/d\phi$ and $H^2(t) = \{ \dot{\phi}(t)^2/2 + V_E[\phi(t)]\}/3$. The initial conditions for the numerical solution of Eq.~(\ref{eqphi}) are the standard ones in the slow-roll approximation \cite{Linde:2007fr}:
$\phi(0) = - \sqrt{2(1 + 2 N)}\approx - 15.5$ and 
$\dot \phi(0) = \sqrt{2/3}\approx 0.81$, 
which correspond to $ R(0)  \approx 
3.4 \times 10^{-3}$ and $\dot R(0) \approx 
1.8 \times 10^{-3}$. \footnote{Where $\phi$ is given in Planck-Mass ($M_{\rm pl}$) units, $R$ is given in $M_{\rm pl}^4$,
and $N=60$ is the number of efolds.} 
We point out t{black}hat in standard $\phi^2$ inflation, the slow roll is an attractor \cite{Grain_2017} so that the initial conditions do not need to be fine tuned. In the corresponding phase in the JF, where we fit $f(R) \approx R^{2.2}$, the same happens. \textcolor{black}{We stress that there is no GR-like term ($\propto R$) in the best-fit curve in this early phase.}

\begin{figure}
\center
\includegraphics[width=0.4\textwidth]{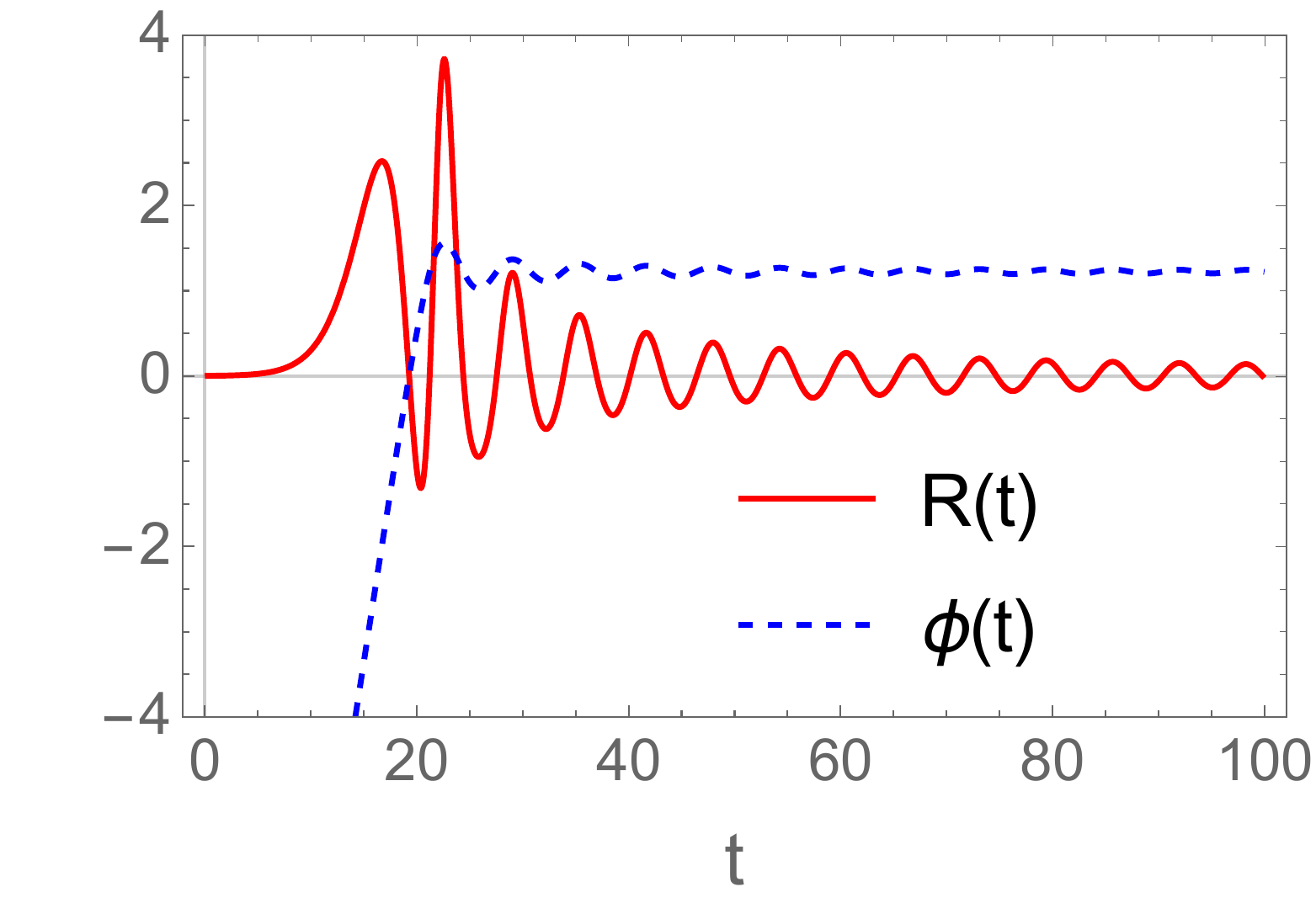}
\caption{Numerical solution for $R(t)$ (red/solid) and $\phi(t)$ (blue/dashed), given by Eq.~(\ref{RVE}) and the numerical solution of Eq.~(\ref{eqphi}), respectively, with $N = 60$ efolds, using the potential defined in Eq.~(\ref{VE}), with $m_\phi = 1$, $\Lambda=0$ and $a=a_*$.}
\label{Rb050}
\end{figure}

One can follow the evolution of the system along the branches in Fig.~\ref{swallowtail} (top left panel): the system starts close to the origin and slowly moves along the first branch (close to the horizontal axis), generating an initially inflationary phase (since $R\approx {\rm const}$). It then quickly sweeps through the second branch (where $f''<0$) and then oscillates around the origin along the almost-linear third branch (where GR is recovered for $a=0$). 

Accordingly, in Fig.~\ref{Ufigs}, the system starts 
 on a stable (asymptotic) solution $\phi\to-\infty$ ($V\to+\infty$), but the slow roll  drives the field towards the origin. Eventually, it settles down at the minimum $\phi=a$ ($V=\exp(-\beta a)$).

The same behavior can be seen in Fig.~\ref{sbin}, as follows: The system  starts at $V\to\infty$, which is a stable configuration only if the temperature is above the binodal curve, i.e, either slightly below or above $T_c$ (thick black curve). On the other hand, if T is low enough (like the lowest gray curve, which corresponds to $T=0$), the system starts at a metastable phase (the binodal region) --- the initial inflationary solution is indeed momentary.
Either way, the effective fluid quickly crosses the spinodal curve (the unstable region) and then oscillates around 
$P=2 T \exp(2 \beta a)$ and $V=\exp(-\beta a)$, indicated by a gray circle for $T=0$ in Fig.~\ref{sbin}. At this temperature, the system ends exactly on the binodal curve. For higher temperatures, though, the system settles down above the binodal line, i.e, in a stable configuration.

Each description above explains the same evolution from a different point of view; each one uses a different  --- but equivalent --- fluid, as we shall see now. 

\subsection{Einstein Frame (EF)}
\begin{figure}[t]
\center
\includegraphics[width=0.4\textwidth]{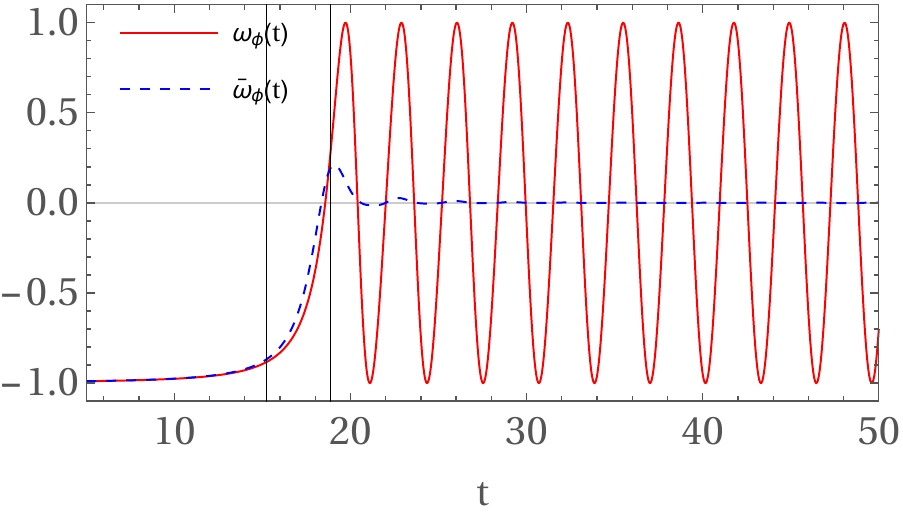}
\caption{Equation-of-state parameter ($w_\phi$ and its time average $\bar{w}_\phi$) for the $\phi$ field, defined in the EF, as functions of time, for $\Lambda=0$. The vertical lines correspond to $t=t_1$ and $t=t_2$, when $\dot R(t_1)=\dot R(t_2)=0$, i.e, at the sideways peaks in Fig.~\ref{swallowtail} (top panels).}
\label{wphi}
 \end{figure}

We plot in Fig.~\ref{wphi}, along each of the aforementioned stages, the corresponding equation-of-state parameter for the $\phi$ field (defined in the EF):
\begin{equation}
    w_\phi(t) \equiv \frac{p_\phi(t)}{\rho_\phi(t)} \equiv 
    \frac{\frac{1}{2}{\dot\phi}^2-V_E[\phi(t)]}{\frac{1}{2}{\dot\phi}^2+V_E[\phi(t)]},
\end{equation}
and its average over one period $T$ (defined in the final oscillatory phase).
There are clearly two distinct phases: the early inflationary period, characterized by $w_\phi \approx \bar{w}_\phi \approx -1$, and the dust-like phase, when $w_\phi$ oscillates between $\pm 1$ and $\bar{w}_\phi=0$, as for the traditional inflaton field in the JF \footnote{At some point, the inflaton field should couple to matter (which is absent in our model from the beginning) to start (p)reheating --- the study of such phase is beyond the scope of the present paper.}. The sideways peaks, at $t_{1,2}$, indicate the transition between the aforementioned phases. 

\subsection{Jordan Frame (JF)}
\begin{figure}[t]
\center
\includegraphics[width=0.4\textwidth]{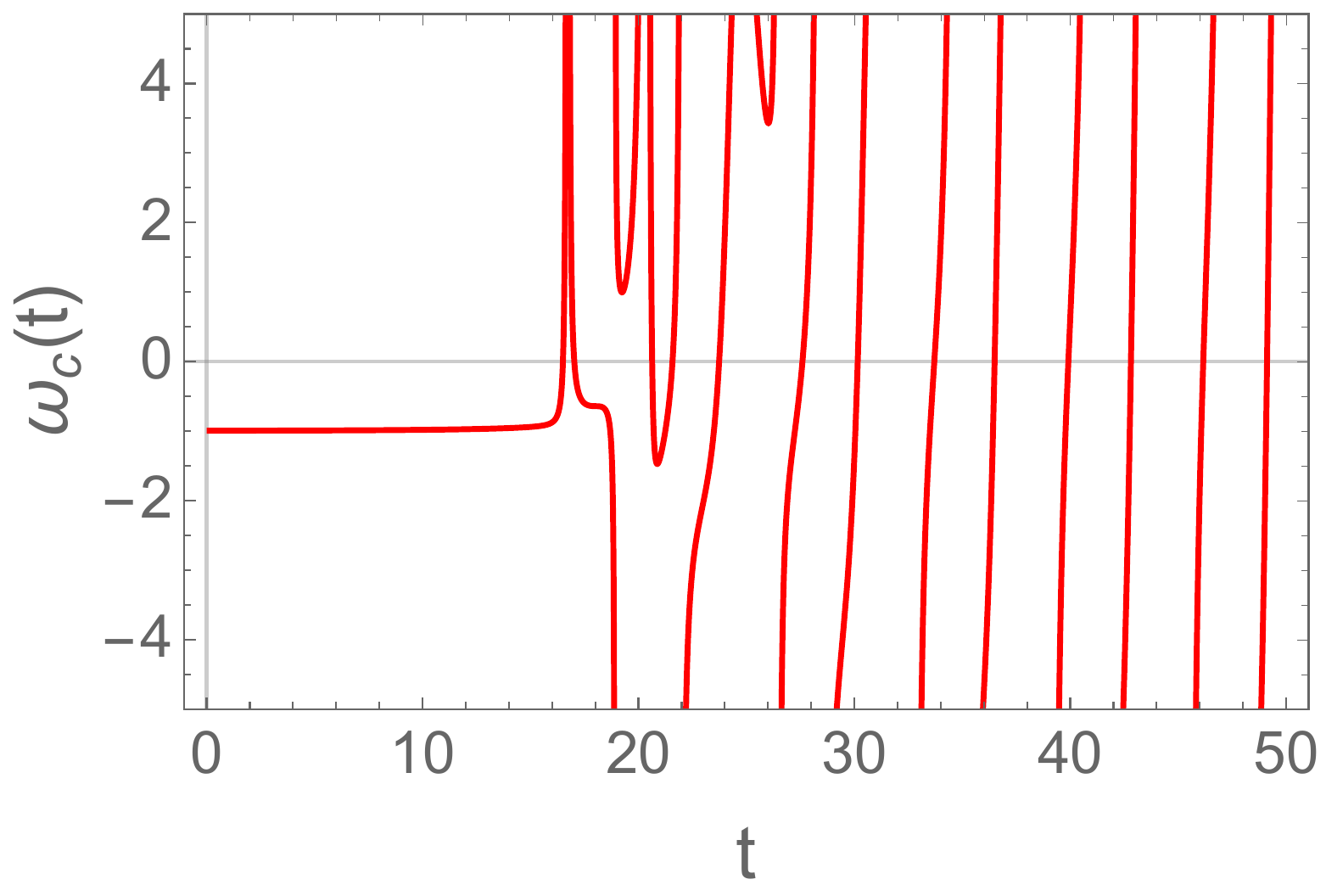}
\includegraphics[width=0.4\textwidth]{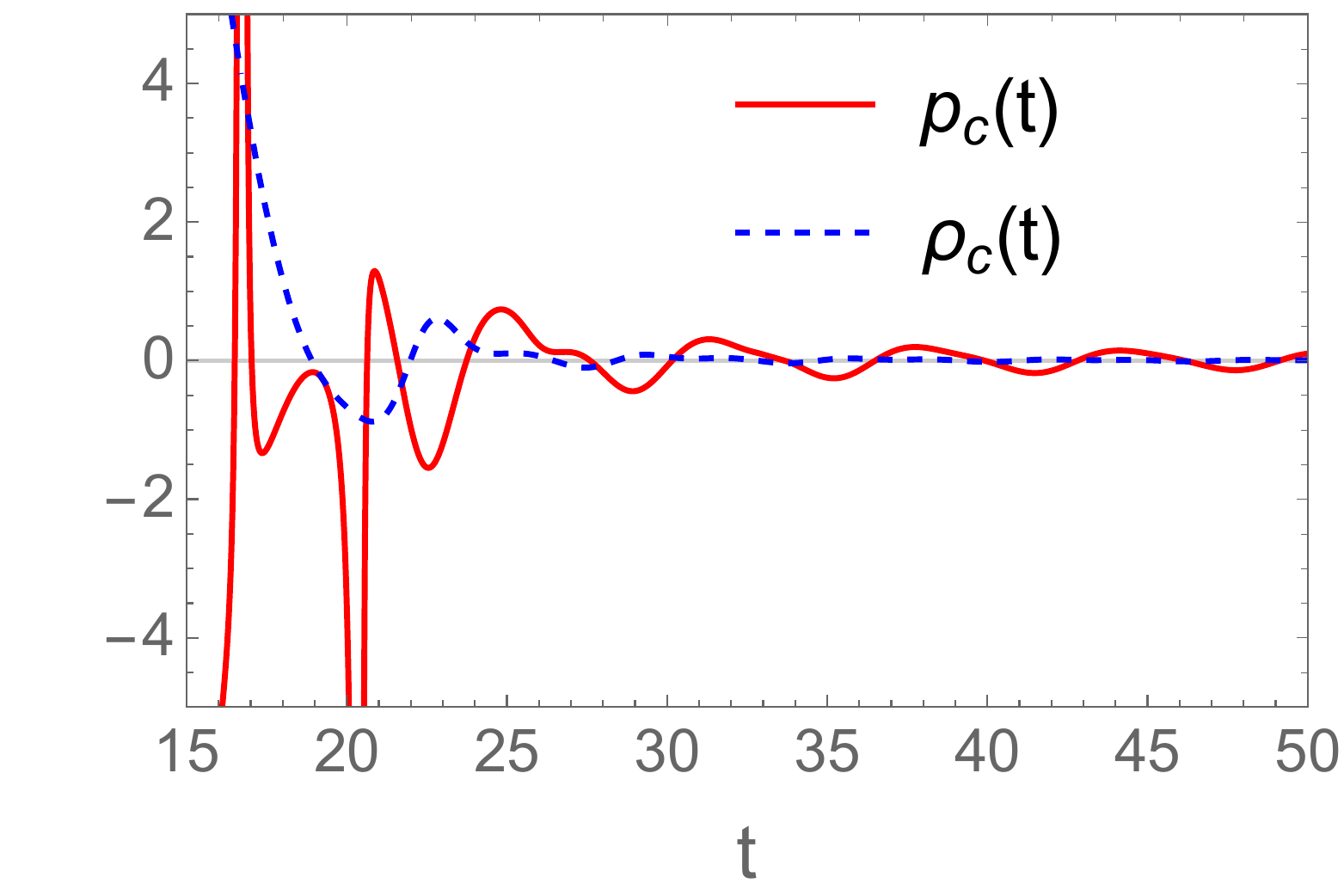}
\caption{{\bf Left panel:} Equation-of-state parameter $\omega_c$ for the ``curvature fluid'' in the JF as a function of time. The divergences, all of them non-physical, correspond to $\rho_c=0$, which happens periodically while the field $\phi$ oscillates around the minimum of its potential $V_E(\phi)$. {\bf Right panel:} Corresponding pressure $p_c$ (red solid curve) and density $\rho_c$ (blue dashed line) for the  ``curvature fluid'', as a function of time. In {\bf both panels}, $\Lambda=0$ and $a=a_*$. 
}
\label{wJF}
\end{figure}

There is a corresponding behavior in the JF, of course. From the extra term in the Einstein equations, one can define a ``curvature fluid'' whose energy density and pressure are, respectively:
\begin{align}
8\pi G \rho _{c} &\equiv \left( f'R - f\right)/2 - 3H\dot{f'} + 3H^{2}(1-f') \label{rhocurv}\\
8\pi G p_{c}  &\equiv \ddot{f'} + 2H\dot{f'} - (2\dot{H}+3H^{2})(1-f') + (f-f' R)/2 \label{pcurv}.
\end{align}
In Fig.~\ref{wJF} we plot the corresponding equation-of-state parameter  $\omega_c \equiv p_c/\rho_c$ (left panel), $\rho_c(t)$, $p_c(t)$ (right panel), all of them defined in the JF, for $\Lambda=0$ and $a=a_*\equiv 1/\beta$. In the inflationary phase, the  curvature fluid behaves as a cosmological constant ($\omega_{c}\approx -1$), as expected, since it is responsible for the accelerated quasi-de Sitter expansion. In the oscillatory phase, on the other hand, the behavior of $\omega_{c}$ diverges just because $\rho_{c}$ vanishes periodically, whenever $\phi(t)=0$ at the bottom of its potential $V_E(\phi)$ --- see Fig.~\ref{wJF}, right-hand panel. Nevertheless, there are no divergences of {\it physical} quantities. 

If $\Lambda\neq 0$, then $\omega_c = \omega_\phi = -1$ also in the final stages, as expected. 

\subsection{vdW fluid}

\begin{figure}[!t]
\center
\includegraphics[width=0.4\textwidth]{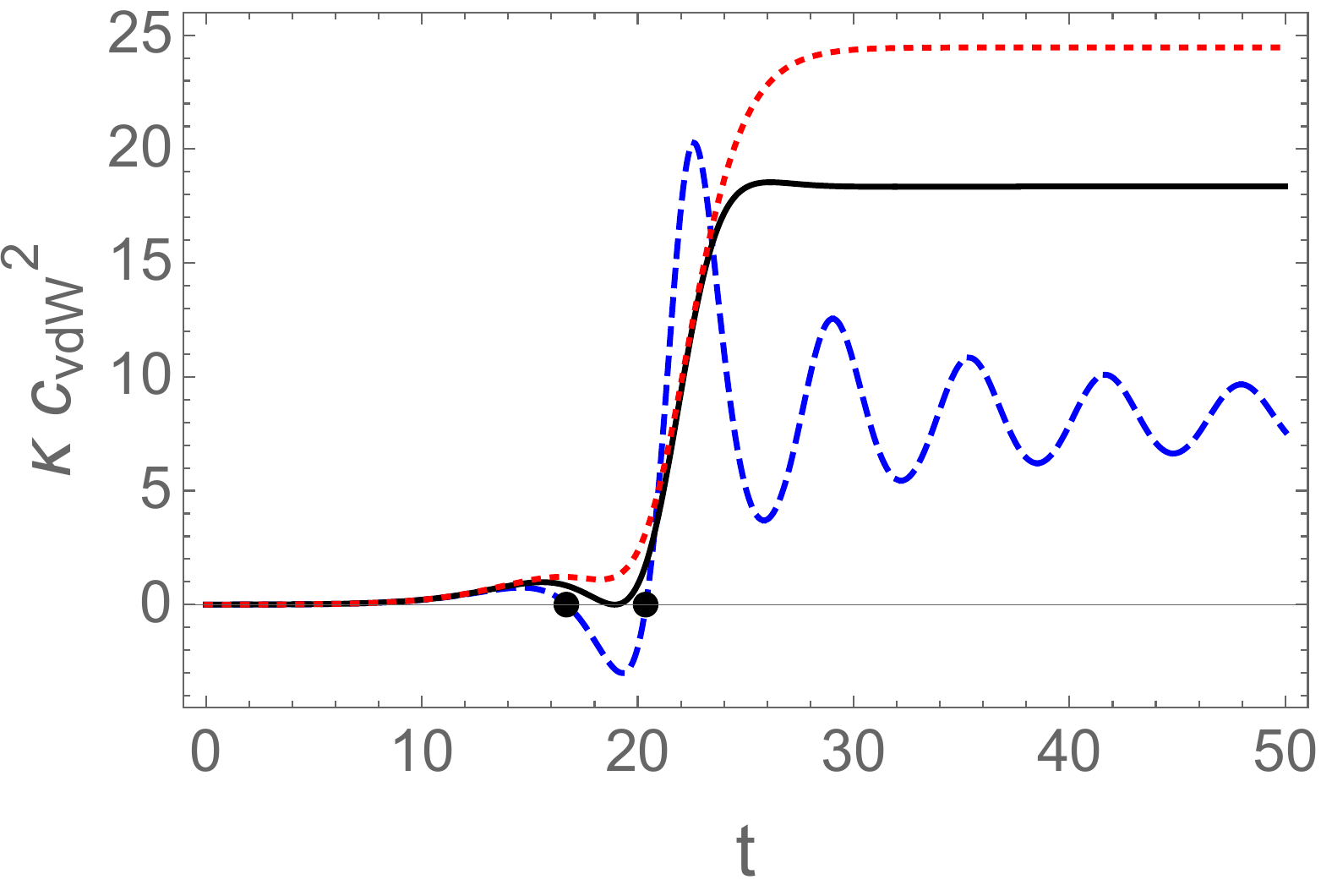}
\caption{Plot of  $\kappa \cdot c_{\rm vdW}^2 \equiv \kappa \cdot \dot P/ \dot \rho$ for the effective vdW gas, for $a=a_*$ and $T=0$ (dashed blue), $T=15/16$ (black) and $T=1.5$ (dotted red), as functions of time. The black dots indicate when $\dot R(t)=0$, i.e, at the sideway peaks in Fig.~\ref{swallowtail}, between which $f''(R)<0$.}
\label{cs2}
\end{figure}

Two important pieces of information are available only from the vdW gas and {\it not} from either the curvature fluid or the $\phi$ field. 

One of them is its sound speed squared, defined as $c_{\rm vdW}^2 \equiv \dot P/\dot \rho = -(V^2/\kappa) \dot P / \dot V$ (where we define $\kappa>0$ by $\rho =: \kappa /V$) and  plotted in Fig.~\ref{cs2}. We can see that $c_{\rm vdW}^2<0$ {\it only} between the first two extrema of $R(t)$, i.e, in the second branch (see Fig.~\ref{swallowtail}), when $f''<0$, as expected from the usual {\it perturbative} argument on stability of $f(R)$ theories \cite{Sotiriou:2008rp}.  Obviously, for $T>T_c$, the second branch is suppressed and one obtains $c_{\rm vdW}^2>0 \, \forall t$. 
With an imaginary sound speed, fluctuations grow exponentially fast, but, during the spinodal decomposition process, only a given range of wavelength do so \cite{CHOMAZ2004263}. This is similar to a feature that has already been proposed in the preheating scenario \cite{preheating}. Further details will be the subject of future work. 

Another important feature is the sudden change in the entropy, from $S(\phi\to-\infty)=0$ to $S(\phi=a)=-2 e^{2 \beta a}$, marking the release of latent heat, just as expected in an ordinary first-order phase transition, which has already been pointed out by the $C_P$ behavior, shown in the previous section. The relation with (p)reheating will also be the subject of future work.

\section{Conclusions}

The toy model here presented is able to generate inflation in the early universe even if $\Lambda=0$, as expected form the standard $\phi^2$ potential in the EF. The mechanism in the JF, on the other hand, is a modification of GR: $f(R)\sim R^{2.2}$, similar to the already known Starobinsky's $R+ \textcolor{black}{\alpha} R^2$ model \textcolor{black}{ \cite{STAROBINSKY198099}  --- see also Ref.~\cite{PhysRevD.91.064016} --- note, however, that those papers keep the traditional GR term ($\propto R$), which is absent here. } 

An unexpected piece of information is brought to light by a third ``frame'', where the system is described by a vdW-like gas. The whole thermodynamics picture then follows: binodal and spinodal curves, phase transition, critical quantities (pressure, volume and temperature), entropy jumps, specific-heat divergence (and the corresponding critical exponent).

\textcolor{black}{Previous works \cite{Motohashi1, Motohashi2} have also found oscillatory behaviour of the effective dark-energy parameter $w_c$ around a (future) de Sitter solution, corresponding to an (perhaps) infinite number of crossings of the phantom divide (when $w_c=-1$).}

We also recall that a non vanishing $a>0$ reduces the value of the effective Newton's constant 
and, at the same time, generates a large effective cosmological constant in the JF (see discussion in Section \ref{conformal}). 
In a more speculative note, we hypothesize that such a mechanism could be used to (almost) cancel out a bare $\Lambda_0$ in the JF, if $\Lambda<0$. Moreover, the cosmological constant in the EF is {\it not} a dynamical quantity in the present work, but it may become so if it is actually the vacuum energy of another field which happens to go through a phase transition of its own.

In any case, the inverse mapping from EF to JF and the phase transition still stand and may be a key feature in a more detailed model. We are currently examining other potentials $V_E(\phi)$ and further generalizations --- Indeed, non-trivial potentials have been investigated before \cite{MIELKE_2008} but with no mention to the thermodynamics we develop here.

SEJ thanks Valerio Faraoni and Thomas Sotiriou, for some clarifying talks in the initial phase of this work, and Robert Brandenberger for important suggestions. CDP thanks Omar Roldan for his kind hospitality, and Diego Restrepo for his support while away from U. Antioquia and acknowledges financial support from COLFUTURO/COLCIENCIAS, Colombia, under the program “Becas Doctorados Nacionales 647” and Instituto de F\'isica from U. Antioquia.
   
%

\end{document}